\documentclass[aps,twocolumn,showpacs,amsmath,amssymb,pra,superscriptaddress,floatfix,longbibliography]{revtex4-1}

\usepackage{bm}
\usepackage{physics}
\usepackage{braket}
\usepackage{amsmath}
\usepackage{amssymb}
\usepackage{mathtools}
\usepackage{graphicx}
\usepackage{dcolumn}
\usepackage{bm}
\usepackage{multirow}
\usepackage{leftidx}
\usepackage{color}
\usepackage{float}
\usepackage{amsfonts}
\usepackage[capitalise]{cleveref}
\usepackage[shortlabels]{enumitem}
\usepackage[german,english]{babel}
\usepackage{gensymb}
\usepackage[thinc]{esdiff}
\graphicspath{ {./Images/Free}, {./Images/bound} }
\setlist[enumerate,1]{label={(\roman*)}}

\begin{document}


\title{Interference in nonlinear Compton scattering using a Schr\"odinger's equation approach}


\author{Akilesh Venkatesh}
\affiliation{Department of Physics and Astronomy, Purdue University, West Lafayette, Indiana 47907, USA}

\author{F. Robicheaux}
\affiliation{Department of Physics and Astronomy, Purdue University, West Lafayette, Indiana 47907, USA}


\date{\today}

\begin{abstract}
The interference between Compton scattering and nonlinear Compton scattering from a two-color field in the X-ray regime is theoretically examined for bound electrons. The underlying phase shifts are analysed using a perturbative approach in the incoming classical field. The perturbative approach is bench marked with a non-perturbative approach in the classical field. The interference for different combinations of linear polarization of the two fields is examined when the Compton and the nonlinear Compton scattered waves have the same wave vector and polarization. Only two cases exhibit interference. When there is interference, the calculations reveal an intrinsic phase difference between the Compton scattered wave function and the nonlinear Compton scattered wave function of either 0 or $\pi$ depending on the scattering angle.
\end{abstract}

\pacs{}

\maketitle

\section{Introduction} \label{Section_introduction}
The advent of high-intensity sources of light has made it possible to probe a wide range of non-linear phenomenon ranging from multi-photon absorption~\cite{multiphoton_ionization,multiphoton_2} to higher harmonic generation~\cite{HHG_review, HHG_1, HHG_theory, HHG_2}. The past two decades saw the first X-ray free electron lasers (XFELs) being commissioned, with some notable ones being the LCLS~\cite{XFEL_SLAC_progressreview, XFEL1, LCLS_5yrs}, SACLA~\cite{SACLA_1,SACLA_2} and the European XFEL~\cite{EuropeanXFEL1, EuropeanXFEL2, EuropeanXFEL3}. In the hard X-ray regime, the LCLS can generate pulses with photon energy between 1-25 keV with a pulse duration of 10-50 fs~\cite{Fuchs,LCLS_specswebsite}. The SACLA facility in Japan can generate pulses of photon energy between 4-20 keV, of pulse duration 2-10 fs~\cite{SACLA_specs_website1, SACLA_specs_website2}. The European XFEL can generate pulses of similar photon energies of up to 25 keV with a pulse duration of $\sim$50 fs~\cite{EuropeanXFEL_chem_specs,EuropeanXFEL1,EuropeanXFEL2,EuropeanXFEL3}. All three facilities can generate laser intensities up to $\sim10^{20}~W/{cm^2}$~\cite{Fuchs, SACLA_specs_website1, SACLA_specs_website2, EuropeanXFEL2, EuropeanXFEL_chem_specs}.  The progress in XFEL technology~\cite{XFEL_DESY_progressreview} in particular has enabled the study of nonlinear Compton scattering. Nonlinear Compton scattering is a term that has been used to refer to several multi-photon scattering processes~\cite{NLC_generalized_defn,NLC_generalized_defn_2}. In this paper, we restrict our discussions of nonlinear Compton scattering to a process where two incoming photons scatter from a free or a bound electron into one outgoing photon. First theoretically described by Brown and Kibble for free electrons in 1964~\cite{KB}, it wasn't until 1996 that it could be experimentally confirmed~\cite{firstNLC1_T}. For an incoming photon of frequency $\omega_{in}$, Brown and Kibble~\cite{KB} showed that the frequency of the nonlinear Compton scattered photon can be obtained approximately using the Compton expression~\cite{Compton}, provided one uses $2\omega_{in}$ for the incoming photon frequency. The scattering angle dependence of the differential cross section for nonlinear Compton scattering~\cite{KB} substantially differs from that of Compton scattering~\cite{KN}.

Despite the emergence of XFELs, experimental analysis of nonlinear Compton scattering has been challenging. One reason for this difficulty is the small size of the nonlinear Compton signal, even with incident field intensities as high as $ \sim 10^{20}~W/cm^2$ ($E=107~a.u.$). In this intensity regime, the nonlinear Compton signal can be six orders of magnitude smaller than the size of the corresponding Compton signal for the same field ~\cite{Krebs, NLCPRA_1}. The relatively few experiments that have studied nonlinear Compton scattering reflects the difficulty. Another major challenge in such an experiment can be the noise from the XFEL itself~\cite{XFEL_theory,XFEL_secondharmonic}. The second harmonic from the XFEL can undergo Compton scattering and add to the noise in the already small nonlinear Compton signal. Both these challenges were discussed in a recent experiment by Fuchs et al.~\cite{Fuchs}.

\begin{figure}
\resizebox{85mm}{!}{\includegraphics{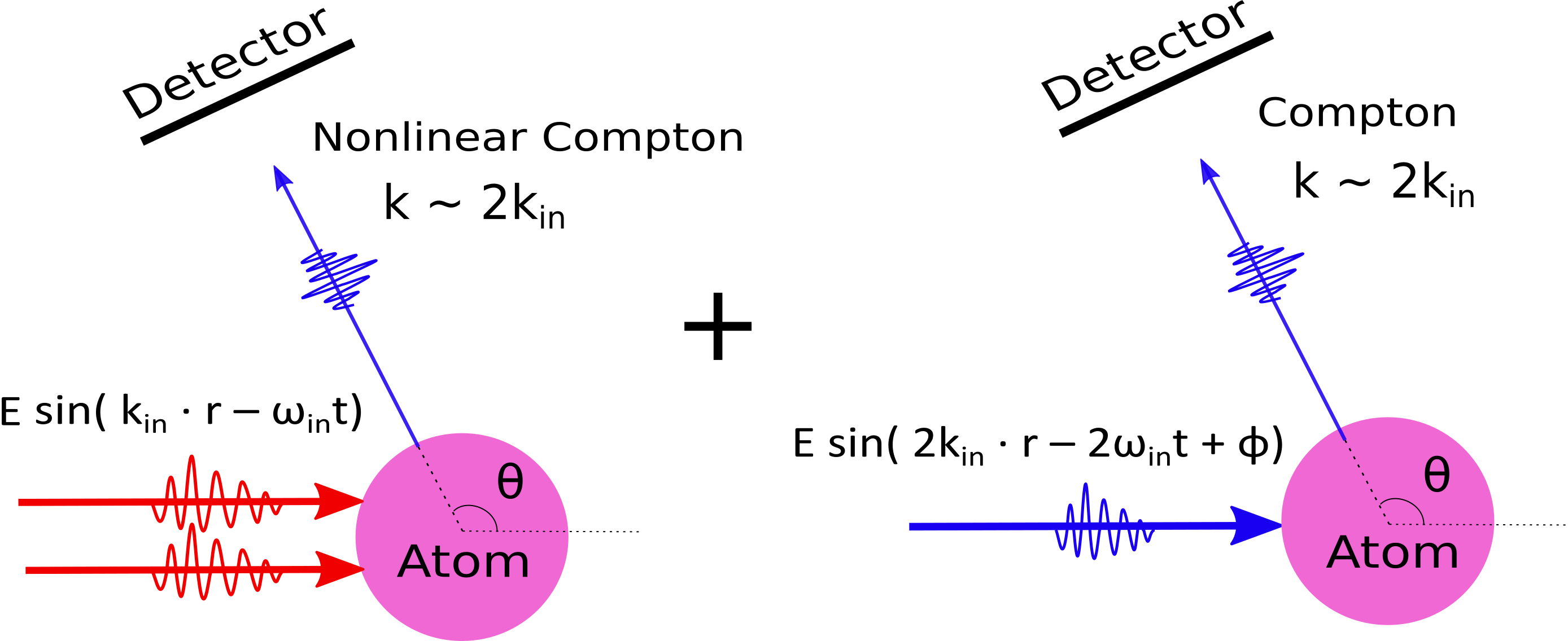}}
\caption{\label{Schematic_diagram} A schematic diagram of the interference between Compton and nonlinear Compton scattering from a bound electron using a two-color field. Here $k_{in}$ refers to the momentum of an incoming photon in the case of nonlinear Compton scattering and $k$ refer to the momentum of an outgoing photon.
}
\end{figure}

In this paper, we study the interference in Compton scattering when using a two-color field of frequency $\omega_{in}$ and $2\omega_{in}$  with a phase difference. The interference is between the Compton scattered photons of the $2\omega_{in}$ field and nonlinear Compton scattered photons of the $\omega_{in}$ field (see Fig.~\ref{Schematic_diagram}). Let the intensities of the $\omega_{in}$ field and the $2 \omega_{in}$ field be $I_{\omega_{in}}$ and $I_{2\omega_{in}}$ respectively. In general, the nonlinear Compton signal scales with the square of the incoming field intensity ($\propto I^2_{\omega_{in}}$) and the Compton signal scales linearly with intensity ($\propto I_{2\omega_{in}}$), for intensities that are within the limits stated in Sec.~\ref{perturbative_vs_exact}. The interference term scales as $\propto {I_{\omega_{in}}}\sqrt{I_{2\omega_{in}}}$. Interference is possible since it cannot be deduced whether the photon came from Compton scattering of the $2\omega_{in}$ field or nonlinear Compton scattering of the $\omega_{in}$ field.

This study suggests techniques to overcome two challenges involved in nonlinear Compton scattering experiments. First, the difference in the intensity between the constructively and the destructively interfered scattered waves, combined with pure Compton scattering measurements can help in determining the extent of nonlinear Compton scattering without having to measure the small signal directly. For example, if the nonlinear Compton signal is 6 orders of magnitude smaller than the Compton signal, then the interference would be ~3 orders of magnitude smaller. Second, the noise from the second-harmonic of the XFEL can be determined by examining the interference between Compton and nonlinear Compton scattering. For this, consider the $\omega_{in}$ field to be the XFEL fundamental. It gives rise to the desired nonlinear Compton signal at $ \sim 2\omega_{in}$ frequency. The second harmonic of the XFEL is the $2\omega_{in}$ field and the Compton scattered photons from this field is the noise at $ \sim 2\omega_{in}$ frequency. Introducing a phase factor ($\phi$) to the $\omega_{in}$ field (or the $2\omega_{in}$ field) and examining the interference can help in identifying the noise.

Several papers in the last few decades, have examined interference effects in multi-photon processes when there occurs an overlap in the initial and final states~\cite{Elliot_interference1,Elliot_interference2,Starace_interference1}. Using two-color fields to analyze interference effects is also not uncommon. For instance, Yin et al.~\cite{Elliot_interference1} examined the interference in the angular distribution of photo-electrons from single and double-photon ionization from a two-color field. Their experiment revealed an interesting asymmetry in the angular distribution despite the initial state of the atom being spherically symmetric.   

The few research works on interference effects involving nonlinear Compton scattering~\cite{NLCinterference_King,NLCinterference_Tobias} have focused on high-energy electrons where, the frequencies of the incident electromagnetic~(EM) waves are around the visible region. These works have also relied on a field-theoretic approach. Unlike the previous work, here we focus on the case of X-ray scattering from bound and non-relativistic free electrons and examine the interference using a Schrodinger equation approach. To understand the interference between Compton and nonlinear Compton scattered wave functions, we use a perturbative approach. We study the dependence of the phase shifts on the frequencies of the incoming field and the binding energy~(BE) of the electron. This analysis is performed over a range of frequencies from 50 a.u. (1.3 keV) to 680 a.u. (18.5 keV). This choice for the frequency is motivated by the typical frequencies accessible from the XFELs~\cite{XFEL_DESY_progressreview,XFEL_SLAC_progressreview} in use and in particular, a recent experiment on nonlinear Compton scattering at the Linac Coherent Light Source at the SLAC National Accelerator Laboratory~\cite{Fuchs}.

This paper is organized as follows: In Sec.~\ref{Methods}, we discuss the theoretical approach to describe Compton and nonlinear Compton scattering both non-perturbatively and perturbatively in the incoming classical field in the limit of non-relativistic electrons. Then, the procedure for studying the interference using them is described. In Sec.~\ref{Results}, the validity of the perturbative approach is demonstrated. Then the case of interference from a two-color field is discussed.

Unless otherwise stated, atomic units will be used throughout this paper.

\section{Methods and Modelling} \label{Methods}
\subsection{Non-perturbative treatment in the classical field} \label{Exact_treatment}
We use a time-dependent Schr\"odinger equation approach to study nonlinear Compton scattering~\cite{Krebs}. The approach is the one described previously~\cite{NLCPRA_1}. This non-relativistic treatment is well justified~\cite{NLCPRA_1} in the regime of nonlinear Compton scattering studied in this paper. This section briefly describes the method; see Ref.~\cite{NLCPRA_1} for a detailed discussion of the derivation. 

The Hamiltonian that describes the laser-electron interaction is given by
\begin{equation}\label{full hamiltonian}
  \hat{H} =  \frac{(\boldsymbol{\hat{P}} + \boldsymbol{\hat{A}})^2}{2} +V(\boldsymbol{\hat{x} }) + \sum_{\boldsymbol{k},\boldsymbol{\epsilon}}\omega_{k}  \hat{a}_{\boldsymbol{k},\boldsymbol{\epsilon}}^{\dagger} \hat{a}_{\boldsymbol{k},\boldsymbol{\epsilon}}
\end{equation}  
where, $\boldsymbol{\hat{P}}$ and $V(\boldsymbol{\hat{x} })$ refer to the momentum operator and the atomic potential experienced by the electron. The quantity $\omega_k$ refers to the angular frequency of the scattered photon. The operators $\hat{a}_{\boldsymbol{k},\boldsymbol{\epsilon}}^\dagger$ and $\hat{a}_{\boldsymbol{k},\boldsymbol{\epsilon}}$ can create or annihilate respectively a photon in the mode $(\boldsymbol{k},\boldsymbol{\epsilon})$. Here $\boldsymbol{k}$ and $\boldsymbol{\epsilon}$ refer to the momentum and the polarization of the scattered photon respectively. The vector potential $\boldsymbol{\hat{A}}$ is written as the sum of the incoming and scattered EM waves. The incoming wave is treated classically while the outgoing wave is quantized~\cite{KB,Loudon}. One can then derive the homogeneous Schr\"odinger equation for the electron in a classical EM-field and the non-homogeneous Schr\"odinger equation for the scattering probability amplitude. The equations are derived by only considering terms up to the first order in the quantized field.  

The homogeneous Schr\"odinger equation describing the wave function of an electron in a classical EM field with no outgoing photons is given by 
\begin{equation} \label{Exact_eq_psi0}
   i \frac{\partial \psi ^ {(0)} }{\partial t} -  \hat{H}_{C}\psi ^{ (0) } = 0
\end{equation}
where,
\begin{equation}
    \hat{H}_{C} = \frac{(\boldsymbol{\hat{P}} + \boldsymbol{A_C})^2}{2}                 + V(\boldsymbol{\hat{x} }).                          
\end{equation}
The quantity $\boldsymbol{A_C}$ refers to the vector potential of the incoming laser pulse. Note that we do not restrict ourselves to the dipole approximation and include the full space and time dependence for the vector potential. Several previous works have examined the effect of dipole approximation and its underlying limitations in the parameter regime studied in this paper~\cite{Stimulatedcompton_PRL,Moe_Forre_ionization}. The explicit space and time dependence ($\boldsymbol{r}$, $t$) is given by, 
\begin{equation}\label{classicalvectorpotential}
\begin{split} 
    \boldsymbol{A}_C = &\frac{E}{\omega_{in}} \cos\bigg[( \omega_{in} t - {\boldsymbol{k_{in}} \cdot \boldsymbol{r} } ) \bigg]  \\
    &\times \exp\Bigg[\frac{-(2 \ln 2) (t - \frac{{\boldsymbol{\hat{k}_{in}} \cdot \boldsymbol{r} }}{c})^2}{t^2_{wid}} \Bigg]  \boldsymbol{\epsilon_{in}}
\end{split}
\end{equation}
where the quantities $E$, $\omega_{in}$, $\boldsymbol{k_{in}}$, $t_{wid}$ and $\boldsymbol{\epsilon_{in}}$ refer to the incoming electric field amplitude, angular frequency, momentum, the full width at half maximum (FWHM) of the pulse intensity and polarization direction respectively. Note that the quantity $\boldsymbol{\hat{k}_{in}}$ refers to a unit vector in the direction of $\boldsymbol{k_{in}}$. 

The non-homogeneous Schr\"odinger equation describes the electron part of the wave function after scattering a photon. It is given by, 
\begin{equation} \label{Exact_eqn_psi1}
\begin{split} 
  i \frac{\partial \psi ^ {(1)}_{\boldsymbol{k},\boldsymbol{\epsilon}} }{\partial t} - \hat{H}_{C} \psi ^{ (1) }_{\boldsymbol{k},\boldsymbol{\epsilon}} =  &\sqrt{\frac{2\pi}{ V\omega_{k}} } e^{-i\boldsymbol{k\cdot r} }  e^{i\omega_{k} t }    \\    
  & \times \boldsymbol{\epsilon}^* \cdot (\boldsymbol{\hat{P}} + \boldsymbol{A}_C ) W(t) \psi ^{(0)}.
\end{split}
\end{equation}
Here, $V$ refers to the quantization volume that comes from quantizing the outgoing field~\cite{Loudon}. The final results for the differential cross section are independent of the quantization volume. The quantity $\psi ^ {(1)}_{\boldsymbol{k},\boldsymbol{\epsilon}}(\boldsymbol{r},t)$ refers to the probability amplitude for a scattered photon to be of momentum $\boldsymbol{k}$ and polarization $\boldsymbol{\epsilon}$ and the electron to be found at position $\boldsymbol{r}$ at time $t$. The quantity W(t) is a smooth windowing function that is used to turn on the source term adiabatically only for the duration of the incoming laser pulse. The reason for W(t) is twofold: First, to prevent the unphysical emission of photons. Second, to find the ground-state of the electron-photon coupled system. Note that the final results are independent of the specific choice of the windowing function as long as W(t)=1 while the classical X-ray field is nonzero and the W(t) turns on and off smoothly enough. 

For the atomic potential, we choose the following: 

\begin{equation} \label{Potential}
     V(r) = \frac{- Z}{2\sqrt{r^2 + a^2} }\big[ {1 + exp(-r)}\big].
 \end{equation}
Here $a$ is a small parameter used to avoid the singularity at the origin. Note that usage of $a^2$ instead of $a$ marks a departure in convention from our previous work~\cite{NLCPRA_1}. A value of $a^2$ = 0.05 a.u. is used for all the calculations in this paper unless otherwise specified. The quantity Z, characterizes the effective nuclear charge, which is varied to model a range of binding energies~(BE) for the electron. This potential was not chosen to reproduce any atomic orbitals but was chosen to give a range of binding energy, confinement distance, and nuclear charge.

These two equations~[Eqs.~(\ref{Exact_eq_psi0}) and (\ref{Exact_eqn_psi1})] are solved numerically in a Cartesian grid of points to obtain the scattering probability~($ P_ {\boldsymbol{k},\boldsymbol{\epsilon}}$)  which is the probability density in k-space for a photon to scatter with momentum $\boldsymbol{k}$ and polarization $\boldsymbol{\epsilon}$. The $ P_ {\boldsymbol{k},\boldsymbol{\epsilon}}$ is defined as,
\begin{equation} \label{scatteringprobability}
     P_ {\boldsymbol{k},\boldsymbol{\epsilon}} = \int_v {\psi_ {\boldsymbol{k},\boldsymbol{\epsilon}}^{(1)}}^* \psi_ {\boldsymbol{k},\boldsymbol{\epsilon}}^{(1)} d^{3}r.
 \end{equation}

\subsection{Perturbative approach in the classical field}\label{perturbative_approach}
To understand the phase shifts involved in the interference between Compton and nonlinear Compton scattering, a perturbative approach in the classical field is used. We begin by expanding the wave function perturbatively in powers of the incoming \emph{classical} field: 

\begin{equation}  \label{series_psi0}
    \psi^{(0)} = \psi_0^{(0)} + \psi_1^{(0)} + \psi_2^{(0)} + ...           
\end{equation}
The subscript refers to the order of the incoming classical field and the superscript refers to the order of the outgoing quantized field. For example, the quantity $\psi_1^{(0)}$  refers to the term that is zeroth order in the quantized field but first order in the classical field.
 
Before we proceed with the perturbative approach, we modify $\hat{H}_{C}$ in the following manner.
\begin{equation} \label{HC_modification}
   \hat{H}_{C} \rightarrow \hat{H}_{C} - \frac{\boldsymbol{A}_C^2}{2}\bigg\rvert_{r=0}
\end{equation}
This is equivalent to adding a pure time-dependent function to the Hamiltonian. Such an addition simply introduces a time-dependent phase factor to the wave function $\psi^{(0)}$. 
\begin{equation} \label{psi0_phasefactor}
   \psi ^ {(0)} \rightarrow e^{i\xi(t)} \psi ^ {(0)}
\end{equation}
where,
\begin{equation} \label{xi_t}
   \xi(t) = \int \frac{\boldsymbol{A}_C^2}{2}\bigg\rvert_{r=0} dt.
\end{equation}
This does not change the physics of $\psi^{(0)}$ as its norm remains the same.

Multiplying Eq.(\ref{Exact_eqn_psi1}) on both sides by $e^{i\xi(t)}$ and taking the phase factor inside the time-derivative yields,
\begin{equation} 
\begin{split}
   i \frac{\partial }{\partial t} \Bigg( e^{i\xi(t)} \psi ^ {(1)}_{\boldsymbol{k},\boldsymbol{\epsilon}}\Bigg) &- \Bigg( \hat{H}_{C} - \dot{\xi}(t) \Bigg) e^{i\xi(t)} \psi ^{ (1) }_{\boldsymbol{k},\boldsymbol{\epsilon}}  \\
   &=  \sqrt{\frac{2\pi}{ V\omega_{k}} } e^{-i\boldsymbol{k\cdot r} }  e^{i\omega_{k} t } \\    
  & \times   \boldsymbol{\epsilon}^* \cdot (\boldsymbol{\hat{P}} + \boldsymbol{A}_C ) W(t) e^{i\xi(t)} \psi ^{(0)}.
\end{split}
\end{equation}
This reveals that the transformation in Eq.~(\ref{HC_modification}) results in $\psi ^{ (1) }_{\boldsymbol{k},\boldsymbol{\epsilon}}$ also gaining the same phase factor $e^{i\xi(t)}$. Therefore, none of the physics associated with $\psi ^{ (1) }_{\boldsymbol{k},\boldsymbol{\epsilon}}$ changes as well. This justifies the transformation in Eq.~(\ref{HC_modification}). By adding this transformation, the norm of the quantity $\psi_0^{(0)}$ + $\psi_1^{(0)}$ +  $\psi_2^{(0)}$ remains $\sim$1 even in cases with a large electric field such as in Fig.~\ref{Fig_breakdown_threshold}.

Along with the transformation in Eq.~(\ref{HC_modification}), one can then substitute Eq.~(\ref{series_psi0}) in the homogeneous Schr\"odinger equation~[Eq.(\ref{Exact_eq_psi0})]. By separating out the terms based on the order of the classical field, the equations for the corresponding wave functions can be derived. The equations for the wave function that is zeroth, first and second order in the classical field respectively are given by:
\begin{equation} \label{homogeneous_zeroth}
   i \frac{\partial \psi_0 ^ {(0)} }{\partial t} -   \hat{H}_a \psi_0 ^{ (0) } = 0
\end{equation}

\begin{equation} \label{homogeneous_first}
   i \frac{\partial \psi_1 ^ {(0)} }{\partial t} -  \hat{H}_a \psi_1 ^{ (0) } = {\boldsymbol{\hat{P}}} \cdot \boldsymbol{A}_C ~ \psi_0 ^ {(0)}
\end{equation}

\begin{equation} \label{homogeneous_second}
\begin{split}
   i \frac{\partial \psi_2 ^ {(0)} }{\partial t} -  \hat{H}_a \psi_2 ^{ (0) } = ~  &(\boldsymbol{\hat{P}}  \cdot \boldsymbol{A}_C) ~ \psi_1 ^ {(0)} \\ 
   &+ \Bigg( \frac{\boldsymbol{A}_C^2}{2} - \frac{\boldsymbol{A}_C^2}{2}\bigg\rvert_{r=0} \Bigg) ~ \psi_0 ^ {(0)}
\end{split}
\end{equation}
where, 
\begin{equation} \label{Hatom}
   \hat{H}_a = \frac{\boldsymbol{\hat{P}}^2} {2}+V(\boldsymbol{\hat{x} }).
\end{equation}
The quantity $\psi_0 ^ {(0)}$ refers to the electronic wave function that does not depend on the external field. $\psi_1 ^ {(0)}$ and $\psi_2 ^ {(0)}$ contain the probability amplitudes for the electron to absorb one and two photons respectively.

Similarly, the first order wave function in the quantized field~$\psi ^ {(1)}_{\boldsymbol{k},\boldsymbol{\epsilon}}$, can also be expanded in a perturbative power series in the classical field:

\begin{equation} \label{series_psi1}
    \psi^{(1)} = \psi_0^{(1)} + \psi_1^{(1)} + \psi_2^{(1)} + ...           
\end{equation}
Note that, the subscripts $\boldsymbol{k},\boldsymbol{\epsilon}$ have been dropped from $\psi^{(1)}$ to reduce clutter in the notation. But one has to keep in mind that every term that is first order in the quantized field will depend on these quantities. 

Using the transformation in Eq.~(\ref{HC_modification}), substituting Eq.(\ref{series_psi1}) in the non-homogeneous Schr\"odinger equation~[Eq.~(\ref{Exact_eqn_psi1})] and separating out the terms based on the order of the classical field yields the following equations for the wave function that is zeroth, first, and second order in the classical field respectively:
\begin{equation} \label{Inhomogeneous_zeroth}
\begin{split} 
  i \frac{\partial \psi_0 ^ {(1)} }{\partial t} - \hat{H}_{a} \psi_0 ^{ (1) } =  &\sqrt{\frac{2\pi}{ V\omega_{k}} } e^{-i\boldsymbol{k\cdot r} }  e^{i\omega_{k} t }    \\    
  & \times \boldsymbol{\epsilon}^* \cdot \boldsymbol{\hat{P}}~ \psi_0^{(0)}~W(t)
\end{split}
\end{equation}

\begin{equation} \label{Inhomogeneous_first}
\begin{split} 
  i \frac{\partial \psi_1 ^ {(1)} }{\partial t} - \hat{H}_{a} \psi_1 ^{ (1) } =  &\sqrt{\frac{2\pi}{ V\omega_{k}} } e^{-i\boldsymbol{k\cdot r} }  e^{i\omega_{k} t }    \\    
  & \times \boldsymbol{\epsilon}^* \cdot ( \boldsymbol{\hat{P}} ~ \psi_1^{(0)} + \boldsymbol{A_C} ~ \psi_0^{(0)} )~W(t) \\
  & + (\boldsymbol{A_C} \cdot \boldsymbol{\hat{P}}) \psi_0^{(1)} 
\end{split}
\end{equation}

\begin{equation} \label{Inhomogeneous_second}
\begin{split} 
  i \frac{\partial \psi_2 ^ {(1)} }{\partial t} - \hat{H}_{a} \psi_2 ^{ (1) } = ~ &  \sqrt{\frac{2\pi}{ V\omega_{k}} } e^{-i\boldsymbol{k\cdot r} }   e^{i\omega_{k} t }    \\    
    & \times \boldsymbol{\epsilon}^* \cdot ( \boldsymbol{\hat{P}} ~ \psi_2^{(0)} + \boldsymbol{A_C} ~ \psi_1^{(0)} )~W(t) \\
    &+ (\boldsymbol{A_C} \cdot \boldsymbol{\hat{P}})~\psi_1^{(1)} \\
    &+ \Bigg( \frac{\boldsymbol{A_C}^2}{2} - \frac{\boldsymbol{A}_C^2}{2}\bigg\rvert_{r=0} \Bigg)~ \psi_0^{(1)} \\
\end{split}
\end{equation}
Here, $\psi_0 ^ {(1)}$ describes the probability amplitude of having one outgoing photon of momentum $\boldsymbol{k}$ and polarization $\boldsymbol{\epsilon}$ when there is no incoming field and the electron to be in position $\boldsymbol{r}$ at time $t$. Similarly, $\psi_1 ^ {(1)}$ is the probability amplitude for the case with one outgoing photon and one incoming photon being absorbed and  $\psi_2 ^ {(1)}$ refers to the case with one outgoing photon but with two incoming photons being absorbed. A detailed analysis of the source terms can be found in Sec.~\ref{subsection_interference_compton_NLC}.

These perturbative equations Eq.~(\ref{homogeneous_zeroth})~-~(\ref{homogeneous_second}) and Eq.~(\ref{Inhomogeneous_zeroth})~-~(\ref{Inhomogeneous_second}) are solved simultaneously using the same numerical framework that was developed in Ref.~\cite{NLCPRA_1} to solve the equations summarized in Sec.~\ref{Exact_treatment}. These perturbative equations can also be solved by first obtaining the Green's function for the atomic system and then using the same Green's function for solving each of the equations with the corresponding source terms.

\subsection{Two-color field} \label{subsection_twocolorfield}
To study the interference, we replace the single incoming laser pulse with two pulses of different frequencies. For the case where one of the incoming pulses has a frequency twice that of the other, the dominant interference pattern would consists of linear Compton photons from the $2\omega_{in}$ field and nonlinear Compton photons from the $\omega_{in}$ field. 

The effect of the two-color pulse is examined both non-perturbatively as well as perturbatively. In the non-perturbative treatment~(Sec.~\ref{Exact_treatment}), this is simulated by simply choosing the incoming vector potential as the resultant of the two vector potentials from each incoming pulse. In the perturbative treatment~(Sec.~\ref{perturbative_approach}), each incoming pulse is treated perturbatively and the results for $\psi^{(1)}$ from each pulse is superposed to obtain the total scattering probability. 

We choose the full width at half maximum of the pulse intensity ($t_{wid}$) of the $2\omega_{in}$ field to be $1/\sqrt{2}$ of that of the $\omega_{in}$ field. This is motivated by a preference for a large overlap for the scattering probability in k-space between Compton and nonlinear Compton scattered photons. The pulse widths are chosen in this manner since it is the second order wave function ($\psi_2^{(1)}$) that matters for nonlinear Compton and $A_C ^2$ effectively would have $1/\sqrt{2}$ of the pulse width of $A_C$.

\subsection{Convergence} \label{convergence}
The amount of convergence is determined in the interference calculations by examining the relative change in the total differential cross section. For the calculations in Fig.~\ref{Fig_breakdown_threshold}, the difference in the differential cross section between a grid size of 24 a.u. and 16 a.u. was under $10^{-3}$\%. The difference in the differential cross section between a grid spacing of 0.1 a.u. and 0.07 a.u. was below 0.8\%. In Fig.~\ref{interference_3401C_135}, for Z=4 and a scattering angle of 135$\degree$, the difference in the calculated differential cross section between a grid size of 24 a.u. and 16 a.u. was under $10^{-9}$\%. The difference in the differential cross section between a grid spacing of 0.1 a.u. and 0.07 a.u. was below 0.08\%. 
For the calculations in Fig.~\ref{Fig_largetwid}, the difference in the differential cross section between a grid size of 30 a.u. and 40 a.u. is under $10^{-9}$\%. The difference in the differential cross section between a grid spacing of 0.1 a.u. and 0.07 a.u. was below 0.09\%. For a discussion on the choice of other parameters, see Sec.~\ref{Results}.

\section{Applications} \label{Results}
\subsection{Perturbative vs non-perturbative} \label{perturbative_vs_exact}

\begin{figure}
\resizebox{80mm}{!}{\includegraphics{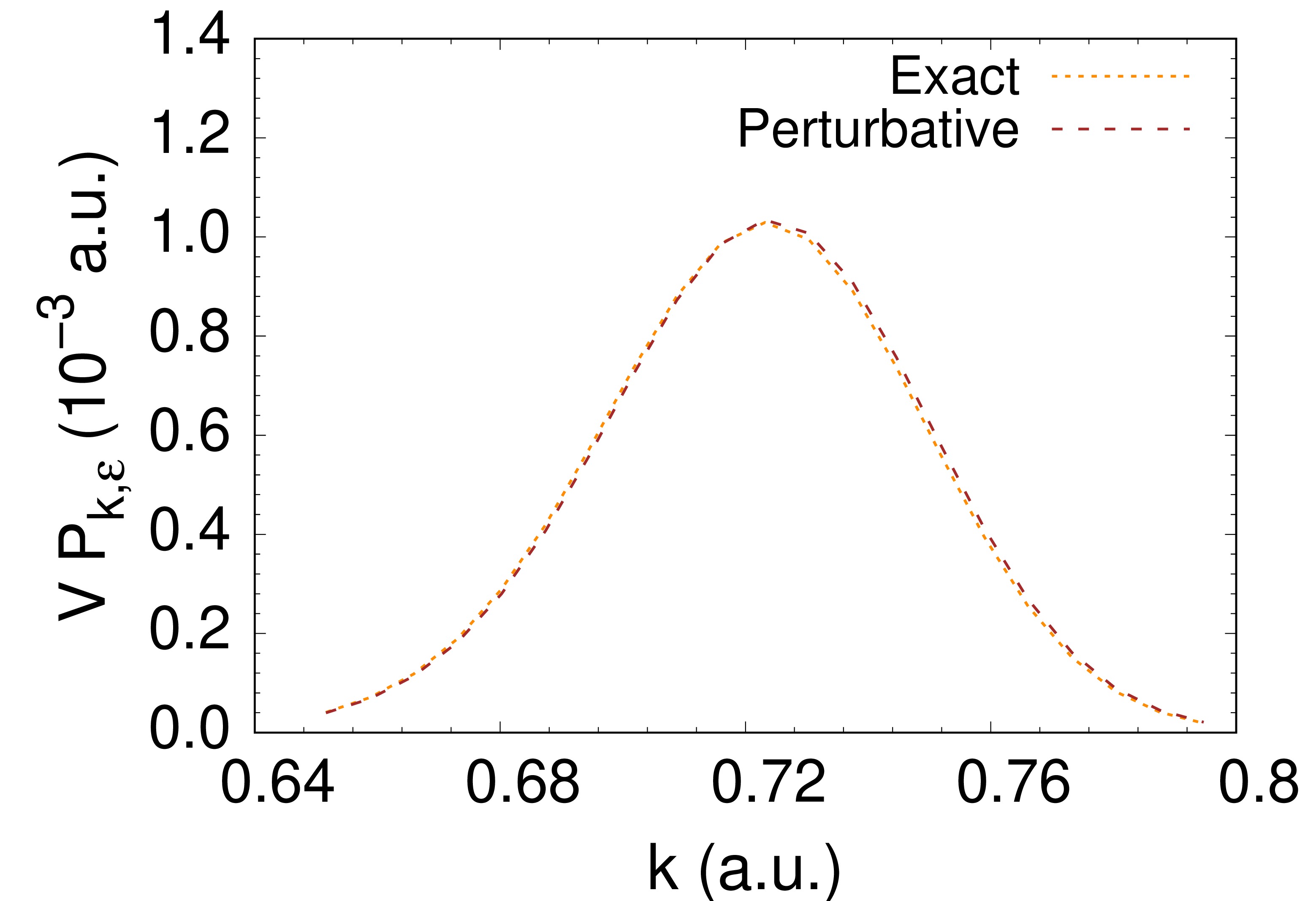}}
\caption{\label{Fig_breakdown_threshold}
Comparison of the results of the non-perturbative treatment and a second-order perturbative calculation in the classical field for the case of nonlinear Compton scattering. The results show a good agreement between the two in the chosen regime. Here $\omega_{in} = 50$ a.u., $E = 502.9$ a.u., $t_{wid} = 0.5$ a.u., $\theta  = 135 \degree$ , $Z = 4$ and $a^2 = 0.05$ a.u. with a binding energy (BE) of 3.306 a.u.
}
\end{figure}


The region of interest involves incoming x-rays with frequencies ($\omega_{in}$) between 50 a.u. and 680 a.u. and electric field amplitudes ($E$) up to a few hundred atomic units which is typical of XFELs in use~\cite{LCLS_5yrs,SACLA_2,EuropeanXFEL3}. In this regime, we find that for the case of linear Compton scattering, a perturbative treatment in the first order in the classical field~[Eqs.~(\ref{homogeneous_zeroth}), (\ref{homogeneous_first}), (\ref{Inhomogeneous_zeroth}) and (\ref{Inhomogeneous_first})] is adequate to describe the problem both for free electrons as well as bound electrons. Nonlinear Compton scattering however, requires a second-order perturbative calculation in the classical field~[Eq.~(\ref{homogeneous_zeroth})~-~(\ref{homogeneous_second}) and Eq.~(\ref{Inhomogeneous_zeroth})~-~(\ref{Inhomogeneous_second})]. 

For the case of nonlinear Compton scattering, the results for the scattering probability obtained from the second order perturbative calculations reveal an excellent agreement with the non-perturbative calculations~(Fig.~\ref{Fig_breakdown_threshold}). Note that the validity of the perturbative expansion depends on the magnitude of $\boldsymbol{A_C}$ ($\sim E/\omega_{in}$). 
For $\omega_{in}$=50 a.u. and electric fields~($E$) below $\sim 600$ a.u. we find that the scattering probability for nonlinear Compton scattering scales with the square of the intensity of the incident wave. As the electric field is increased beyond $E \sim 600$ a.u. (with the other parameters fixed), the scattering probability starts exhibiting non-perturbative behavior. Note that in this regime, Compton scattering adheres to first order perturbative behavior in the classical field as well as the quantized field.

It might be of interest to observe that while the Keldysh paramater in this regime (Fig.~\ref{Fig_breakdown_threshold}) might be less than 1. Tunnel ionization is not expected to be significant because the frequency of the laser is much greater than the classical orbital frequency of the electron~\cite{Keldyshparamter_dichotomy}. Also, when tunnel ionization becomes significant, the framework described in Sec.~\ref{Methods} is adequate since, it includes the full spatial dependence of the incoming laser field and the potential of the electron.

A recent study of photo-absorption probabilities~\cite{Moe_Forre_ionization} in this parameter regime~(Fig.~\ref{Fig_breakdown_threshold}) exhibited non-perturbative behaviour for one-photon net absorption. In the framework described in this paper, the complete information about photo-absorption is contained in $\psi^{(0)}$ [Eq.(\ref{Exact_eq_psi0})]. Calculations of ionization probabilities from $\psi^{(0)}$ reveals that the one-photon net absorption probability does exhibit non-perturbative behaviour in this regime in agreement with Ref.~\cite{Moe_Forre_ionization}. However, the nonlinear Compton scattering probability exhibits perturbative behaviour (Fig.~\ref{Fig_breakdown_threshold}). 

\subsection{Interference between Compton and nonlinear Compton}\label{subsection_interference_compton_NLC}
We now examine the interference effect in the scattered photons when the incoming field consist of two different frequencies with one being twice that of the other and with a phase shift ($\phi$) imposed on the $2\omega_{in}$ field.

To understand how the phase difference in the incoming field affects the scattering probability, we use the perturbative framework developed in Sec. \ref{perturbative_approach}. The interference between the Compton scattered photons and nonlinear Compton scattered photons from the two incoming fields can be understood as the superposition of the scattering probability amplitudes from each field alone. For Compton scattering from the $2\omega_{in}$ field, a first order perturbative calculation in the classical field is used to obtain the scattering probability amplitude. For nonlinear Compton scattering from the $\omega_{in}$ field, a second order perturbative calculation is needed. 

The resultant scattering probability amplitude leading to photons with momentum $\sim 2 k_{in}$ is given by,

\begin{equation} \label{interference_totalwfn}
    \psi_{total}^{(1)} =  {\psi}_{1,2\omega_{in}}^{(1)}(\phi) + {\psi}_{2,\omega_{in}}^{(1)}
\end{equation}
where the first term on the right-hand side (${\psi}_{1,2\omega_{in}}^{(1)}(\phi)$) is the Compton scattering probability amplitude from the $2\omega_{in}$ field which is first order in the classical field. The second term (${\psi}_{2,\omega_{in}}^{(1)}$ ) is the nonlinear Compton scattering probability amplitude from the $\omega_{in}$ field that is second order in the classical field. Here $\psi_{1,2\omega_{in}}$ depends on the phase shift $\phi$. Also, only the two terms that are in Eq.~(\ref{interference_totalwfn}) are relevant because the frequency bandwidth of the incoming field is small compared to the frequency $\omega_{in}$ so that the peaks in the scattering probabilities [Eq.~(\ref{scatteringprobability})] are localized in k-space.

In Eq.~(\ref{interference_totalwfn}), the phase dependence of ${\psi}_{1,2\omega_{in}}^{(1)}(\phi)$ can be determined by examining the three source terms in the first order non-homogeneous differential equation [Eq.~(\ref{Inhomogeneous_first}) ]. The first source term (S1) determined by $\boldsymbol{\hat{P}} ~ \psi_1^{(0)}$, is due to photo-absorption, the second source term (S2) determined by $\boldsymbol{A_C} ~ \psi_0^{(0)}$, is due to pure Compton scattering, the third (S3) determined by $\boldsymbol{A_C} \cdot \boldsymbol{\hat{P}} \psi_0^{(1)}$, describes the laser-dressing of virtual photon emission. The third term (S3) can also be thought of as emission of a final state photon followed by absorption of an incoming photon. Some communities refer to this as u-channel Compton scattering. Here, note that all three terms depend on $\boldsymbol{A_C}$. The $\boldsymbol{A_C}$ being real contains terms of the form $e^{i(2k_{in} \cdot r - 2\omega_{in} t + \phi) }$ and $e^{-i(2k_{in} \cdot r - 2\omega_{in} t + \phi)}$, where $2\omega_{in}$ and $2k_{in}$ refers to the angular frequency and momentum respectively of the incoming field. One can then employ the rotating wave approximation, which would result in only the term with $e^{i(2k_{in} \cdot r - 2\omega_{in} t + \phi)}$ surviving. This leads to the phase factor ($e^{i\phi}$) appearing in every source term (S1, S2 and S3) and hence the final wave function~($\psi_1 ^ {(1)}$) for that field. Therefore Eq.~(\ref{interference_totalwfn}) becomes,

\begin{equation} \label{totalwfn_interference1C_2C}
    \psi_{total}^{(1)} = {\psi}_{1,2\omega_{in}}^{(1)}(0)~e^{i \phi} + {\psi}_{2,\omega_{in}}^{(1)} 
\end{equation}
where ${\psi}_{1,2\omega_{in}}^{(1)}(0)$ is calculated at $\phi=0$.

The contribution from the three source terms to ${\psi}_{1,2\omega_{in}}^{(1)}$ are not of the same size. Given that the incoming EM waves are in the X-ray regime, the photo-absorption term (S1) is small with respect to the Compton term (S2)~\cite{Source_greencomment,Greene_comment}. Also, the contribution from the terms S1 and S3 appear to be of comparable size to each other, but they appear to have a phase factor between them. Their combined scattering probability amplitude is found to be an order of magnitude lower than each of them individually. 

As an added check on the approximations made so far, we compared our results from both the non-perturbative and perturbative approaches and they showed an excellent agreement within the perturbative regime.

\begin{figure}
\resizebox{80mm}{!}{\includegraphics{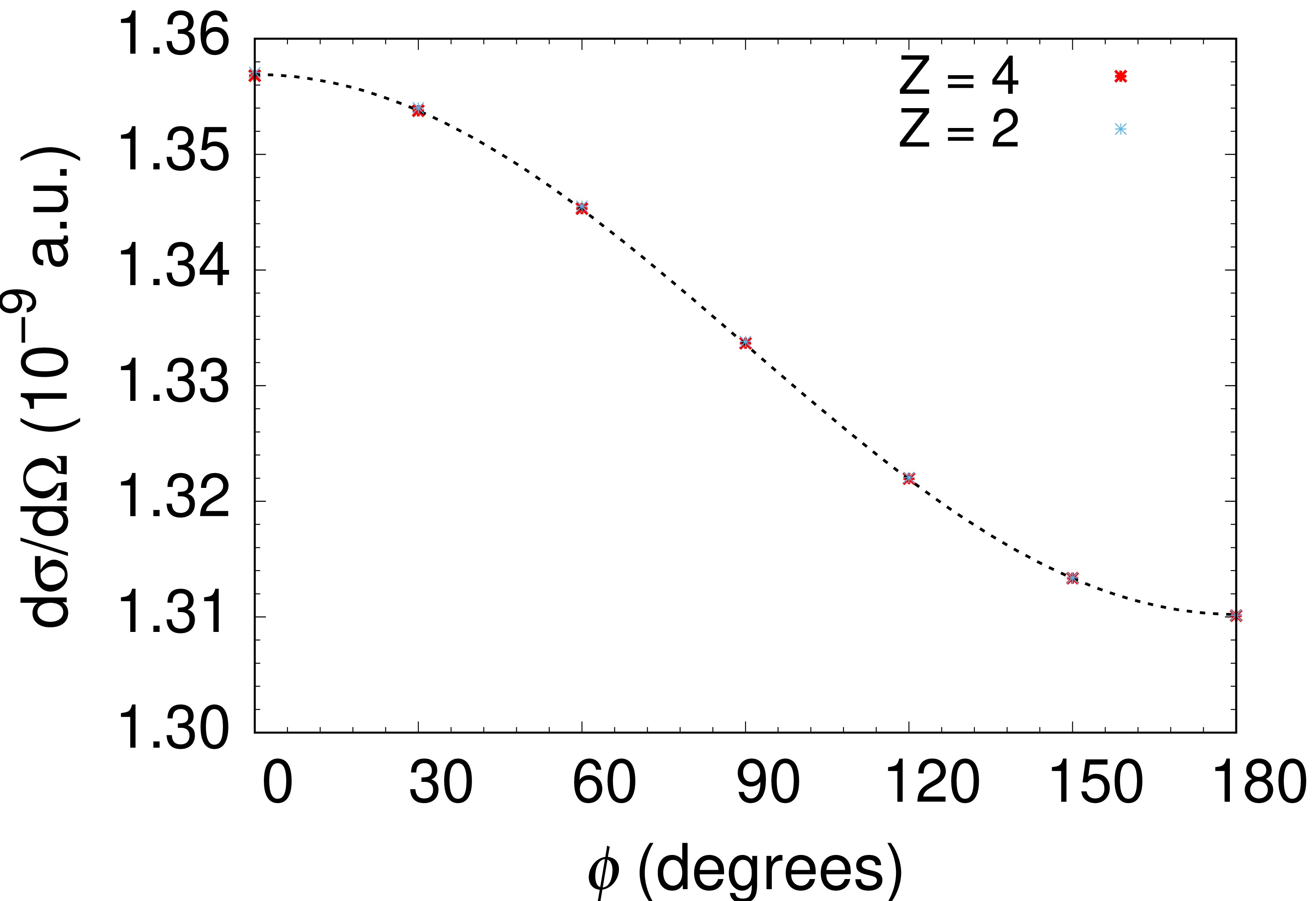}}
\caption{\label{interference_3401C_135}
The figure shows the differential cross section computed using the interfered wave functions from Compton and nonlinear Compton scattering as a function of the imposed phase difference $\phi$ on the $2\omega_{in}$ field for a scattering angle of 135$\degree$. The dotted line is a curve fit of the form C + D cos $\phi$. The plot reveals that there is no intrinsic phase difference between the Compton and nonlinear Compton scattered wave functions. It is clear, there is almost no effect of the binding energy (BE) on the interference pattern in the chosen parameter regime. Here $\omega_{in}$ = 170 a.u. , E = 107 a.u. , $t_{wid}$ = 0.1 a.u. for the $\omega_{in}$ field. Both initial and final polarizations are in the scattering plane. The BE for $Z$ = 2 and $Z$ = 4 are 0.8744 a.u. and 3.306 a.u. respectively. The parameter $a$ remains the same for both with $a^2$ = 0.05 a.u.
}
\end{figure}

\begin{figure}
\resizebox{80mm}{!}{\includegraphics{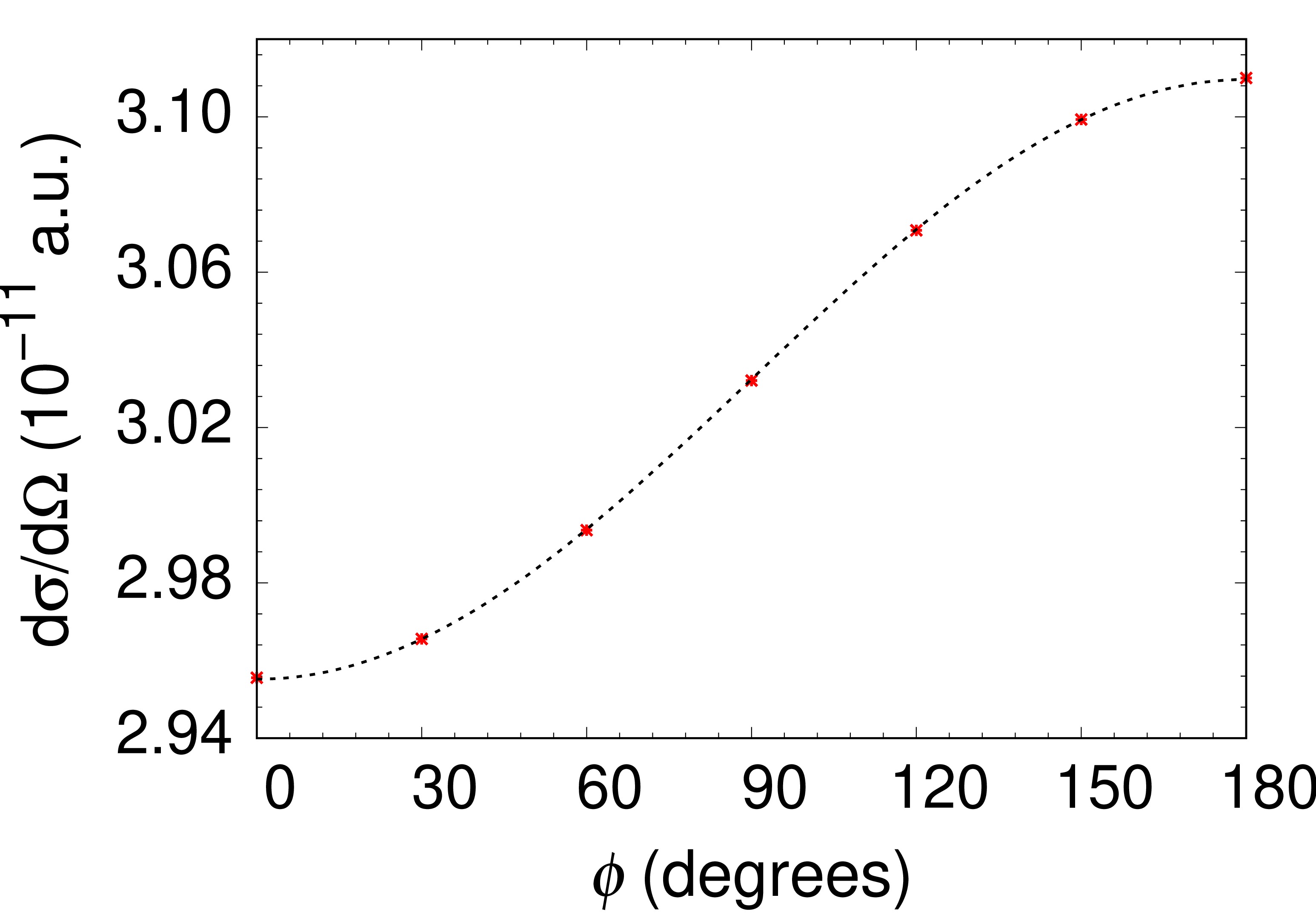}}
\caption{\label{interference_3401C_84}
The figure shows the differential cross section computed using the interfered wave functions from Compton and nonlinear Compton scattering as a function of the imposed phase difference $\phi$ on the $2\omega_{in}$ field for a scattering angle of 84$\degree$. The dotted line is curve fit of the form C + D cos $\phi$. The plot reveals that there is a intrinsic phase difference of $\pi$ between the Compton and nonlinear Compton scattered wave functions. Here $Z$= 4 with the other parameters remaining the same as in Fig.~\ref{interference_3401C_135}.
}
\end{figure}

\begin{figure}
\resizebox{80mm}{!}{\includegraphics{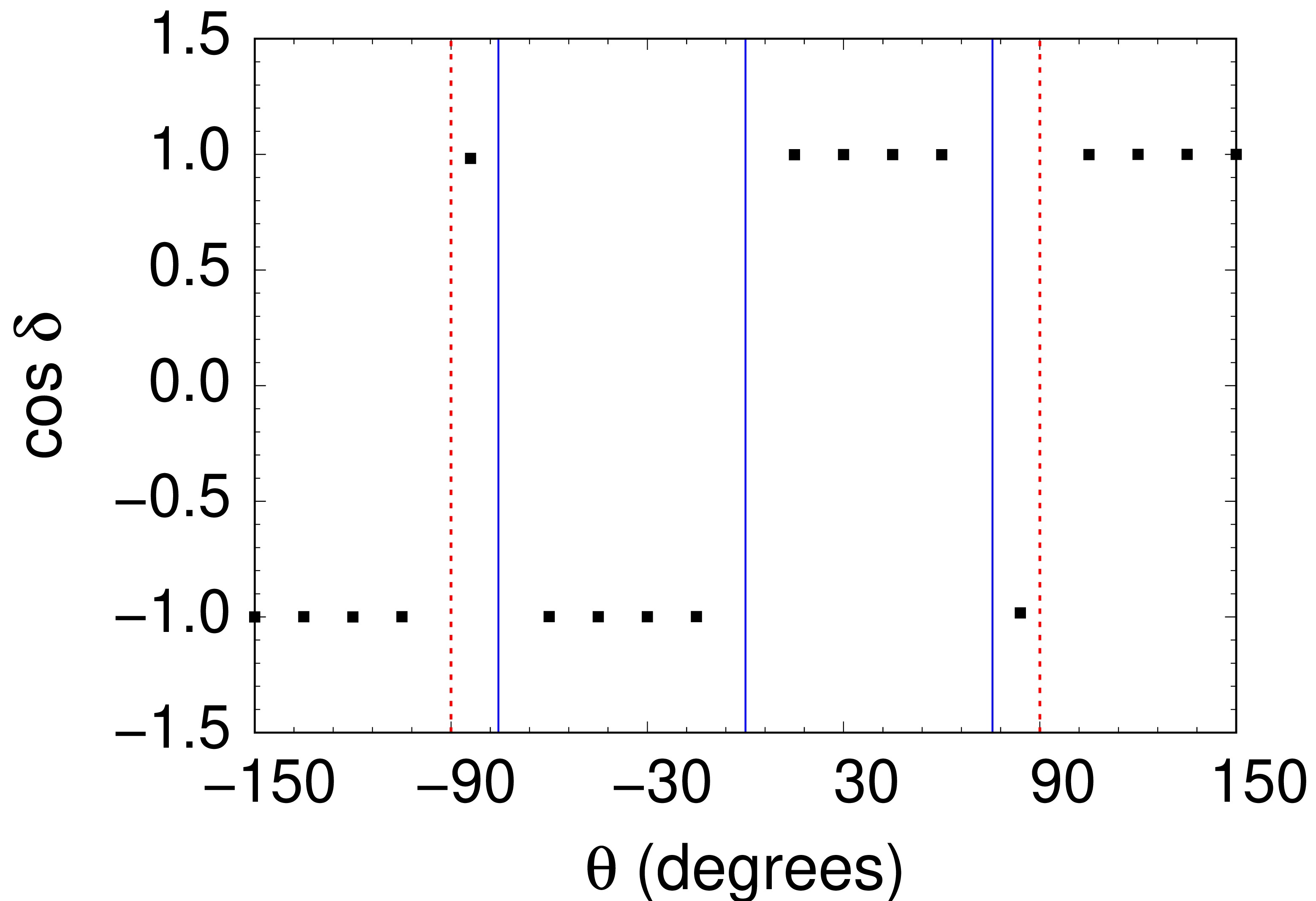}}
\caption{\label{Bnormalized}
The figure shows the dependence of the intrinsic phase difference $\delta$ (black dots) versus the scattering angle $\theta$. The blue solid line indicates the zeroes in the differential cross section of Brown and Kibble~\cite{KB}. The red dotted line indicates the zeroes in the differential cross section of Compton scattering. The calculation reveals a discontinuous jump in the intrinsic phase difference ($\delta$) at scattering angles which are zeroes of the differential cross section for Compton or nonlinear Compton scattering~\cite{KB}. Here $Z$= 4 with the other parameters remaining the same as in Fig.~\ref{interference_3401C_135}
}
\end{figure}

From the total scattered wave function, $\psi_{total}^{(1)}$, the total scattering probability $P_{Tot}$ can be obtained.

\begin{equation} \label{totalprob_interference1C_2C}
\begin{split}
P_{Tot} = & \int_v{|{\psi}_{1,2\omega_{in}}^{(1)}|^2 ~ d^{3}r} +  \int_v{|\psi_{2,\omega_{in}}^{(1)}|^2 ~ d^{3}r } \\ 
&+ \int_v{\big( e^{-i \phi} {\psi}_{1,2\omega_{in}}^{(1)~*} ~ \psi_{2,\omega_{in}}^{(1)} }  \\
&+ e^{i\phi}  {\psi}_{1,2\omega_{in}}^{(1)} ~ \psi_{2,\omega_{in}}^{(1)~*} ~ \big)~d^{3}r 
\end{split}
\end{equation}
Of the 4 terms on the right-hand side, the first term represents the Compton scattering probability and scales linearly with intensity ($\propto I_{2\omega_{in}}$). The second term represents the nonlinear Compton scattering probability and scales quadratically with the intensity($\propto I_{\omega_{in}}^2$). The third and fourth terms together gives rise to the interference. Both the third and the fourth terms are $\propto I_{\omega_{in}} \sqrt{I_{2\omega_{in}}}$. To illustrate this dependence, consider the case when scattering probability for nonlinear Compton is 1\% of that of Compton. Then, the interference term can be as large as $\sim$20\% of the Compton scattering probability.

We choose the two incoming fields to be of equal electric field amplitude with $E=107$ a.u., with polarizations in the same direction and frequency $\omega_{in}$ = 170 a.u. The scattered photon momentum and its polarization are both chosen to be in the same plane as the incoming fields and a range of scattering angles ($\theta$) from 0$\degree$ to 180$\degree$ are considered. Other cases for these quantities are explored after that.

We evaluate the differential cross section from the total scattering probability [Eq.~(\ref{totalprob_interference1C_2C})] using the expression for one-photon differential cross section from Ref.~\cite{NLCPRA_1}. The differential cross section is then given by,
\begin{equation} \label{diffcross_compton}
        \dv{\sigma}{\Omega}^{(1)} =
        \frac{2 V \omega_{in} }{(2\pi)^3}\frac{\int \sum\limits_{ \boldsymbol{\epsilon} } P_ {Tot} k^2dk }{\int I_{2\omega_{in}} dt }.
\end{equation}
Note that we use the Compton differential cross section expression even though the total scattered wave function is the result of interference between Compton and nonlinear Compton. For the intensities chosen ($E = 107$ a.u. for both), this is reasonable since the Compton scattering probability is at least 3 orders of magnitude more than the nonlinear Compton scattering probability. Therefore, in the case of interference between linear and nonlinear Compton scattering, from Eqs.~(\ref{totalprob_interference1C_2C}) and (\ref{diffcross_compton}),  we expect the differential cross section to be of the following form:
\begin{equation} \label{expectation_interference_CD}
     \dv{\sigma}{\Omega} = C + D~cos (\phi - \delta).
\end{equation}
Here C is the Klein-Nishina differential cross section for Compton scattering when the scattering probability for non-linear Compton is much smaller than that of linear Compton. D arises from the interference term in Eq.~(\ref{totalprob_interference1C_2C}) and is proportional to $I_{\omega_{in}} / \sqrt{I_{2\omega_{in} } }$. The quantity $\delta$ is the intrinsic phase shift between the probability amplitude of Compton and nonlinear Compton scattering. 

As a reminder, for the case of free-electrons, the differential cross section for Compton scattering is given by the Klein-Nishina formula~\cite{KN} and for nonlinear Compton scattering is described by Brown and Kibble~\cite{KB}. These expressions are specified below in Eq.~(\ref{Kleinnishinaformula}) and Eq.~(\ref{brownkibbleformula}).
\begin{equation} \label{Kleinnishinaformula}
    \dv{\sigma^{(1)}}{\Omega} = \frac{r^2_e}{2} \bigg( \frac{\omega_k}{\omega_{in}}\bigg)^2 \bigg[ \frac{\omega_k}{\omega_{in}} + \frac{\omega_{in}}{\omega_k} - 2 (\hat{\boldsymbol{k}}_{in} \cdot \boldsymbol{\epsilon})^2 \bigg] 
\end{equation}
where $r_e$ denotes the classical electron radius~\cite{stevenweinberg_book}.
The non-relativistic expression for the differential cross section of nonlinear Compton scattering by Brown and Kibble~\cite{KB} is,
\begin{equation} \label{brownkibbleformula}
    \dv{\sigma^{(2)}}{\Omega} = (\nu r_e)^2 \bigg[ 2(\boldsymbol{\epsilon}_{in} \cdot \boldsymbol{\epsilon})(\hat{\boldsymbol{k}} \cdot \boldsymbol{\epsilon}_{in}) + \frac{1}{2}(\hat{\boldsymbol{k}}_{in} \cdot \epsilon) \bigg]^2
\end{equation}
where,
\begin{equation}
    \nu = (\frac{\alpha E}{\sqrt{2} \omega_{in}})
\end{equation}
and $\alpha$ is the fine structure constant and is equal to $1/c$ in atomic units with c being speed of light in vacuum.

The results of the calculation are shown in Fig.~\ref{interference_3401C_135} ($\theta = 135\degree$) and Fig.~\ref{interference_3401C_84} ($\theta = 84 \degree$). The results are consistent with the dependence expected [Eq.~(\ref{expectation_interference_CD})]. The plot (Fig.~\ref{interference_3401C_135}) shows that the intrinsic phase difference ($\delta$) between the scattering probability amplitude of Compton and nonlinear Compton to be zero for $\theta$ = 135$\degree$. Further investigation reveals that the intrinsic phase difference $\delta$ depends on the scattering angle~$\theta$ (see Fig.~\ref{Bnormalized}). It is zero for scattering angles between $\theta$ = 0 and  $\theta \sim 75\degree$ which is a zero of the nonlinear Compton differential cross section. If the scattering angle is increased beyond this value ($\theta \sim 75 \degree$), the intrinsic phase difference ($\delta$) jumps to a value of $\pi$ (Fig.~\ref{interference_3401C_84}) and it drops back to zero if you increase $\theta$ beyond $90\degree$. It is evident that the intrinsic phase difference switches between a value of 0 or $\pi$ every time the scattering angle crosses a zero of the differential cross section of Compton or nonlinear Compton scattering~\cite{KB}. This is confirmed if one chooses a negative scattering angle. For $\theta = -30 \degree$, the intrinsic phase difference is $\pi$ because there lies a zero of the differential cross section for nonlinear Compton scattering at $\theta = 0\degree$. Note that the Compton scattering differential cross section [Eq.~(\ref{Kleinnishinaformula})] given by the Klein-Nishina formula \emph{does not have a zero} but rather a minimum at $\theta \sim 90 \degree$. But, for our non-relativistic calculation the differential cross section goes to zero in the absence of the Compton profile.

A comparison of the scattering angle dependence of C and the Klein-Nishina formula gave a good agreement with the difference between them being under 0.3\%. Also it is evident from Fig.~\ref{interference_3401C_135}, that the binding energy(BE) of the ground state of the electron does not have a significant effect on the interference between Compton and nonlinear Compton scattering. The difference in the differential cross section between Z = 2 and Z = 4 calculation(Fig.~\ref{interference_3401C_135}) is under 0.02 \%.  

\begin{figure}
\resizebox{80mm}{!}{\includegraphics{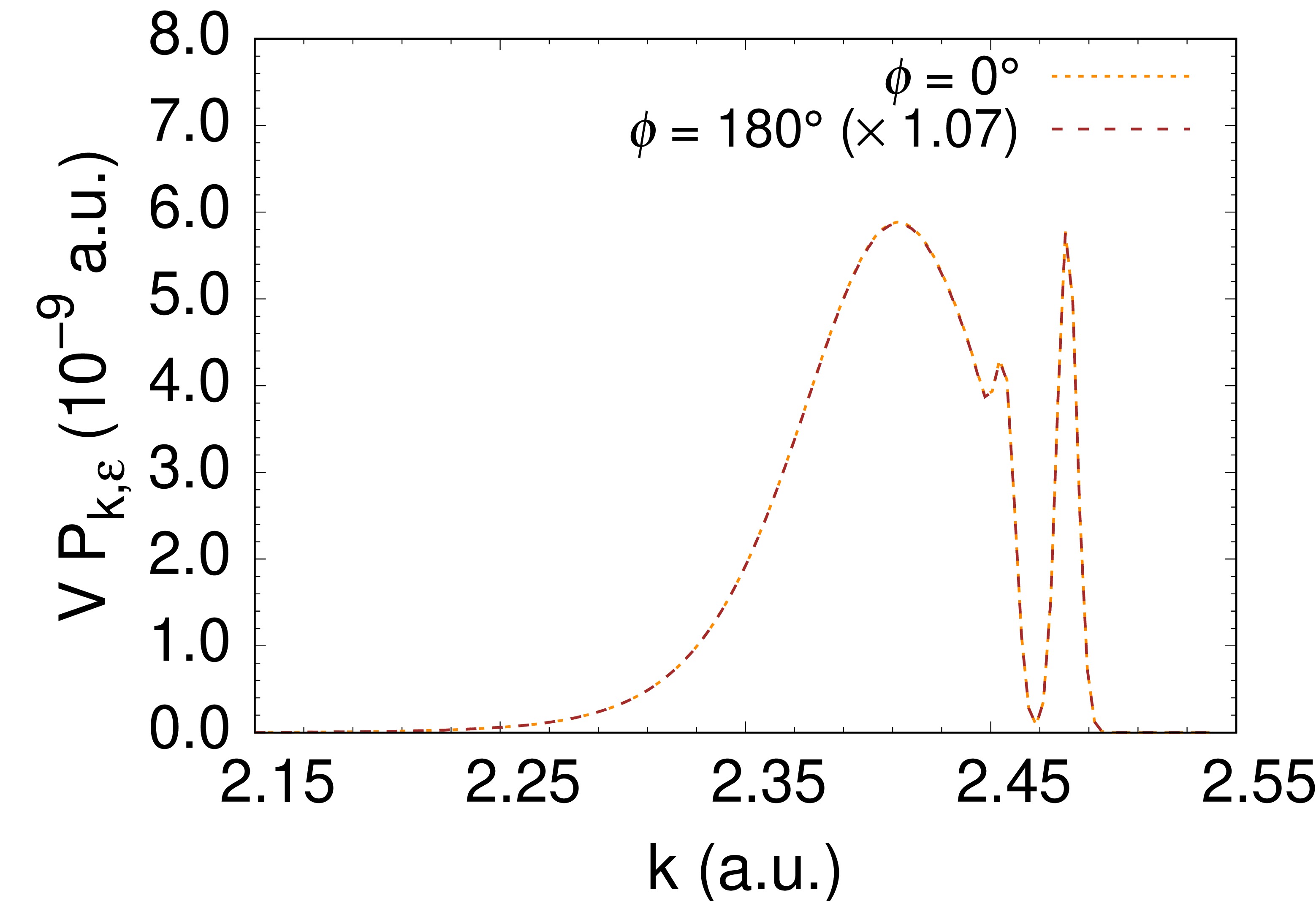}}
\caption{\label{Fig_largetwid}
The results of a 2D calculation for the total scattering probability as a function of scattered photon momentum for the interference between Compton and nonlinear Compton scattering from a two-color field. The two curves correspond to the cases when the imposed phase difference ($\phi$) is $0 \degree$ (constructive) and  $180 \degree$ (destructive). The curve corresponding to $\phi = 180 \degree$ has been scaled by a constant factor to make it coincide with the $\phi = 0 \degree$ curve at the first peak. The first peak ($k\sim 2.41$ a.u.) describes the case of inelastic scattering of the incoming photons from the electron and the second peak ($k\sim 2.48$ a.u.) describes elastic scattering. The coincidence of the two curves ( $\phi=0 \degree$ and $\phi=180\degree$) reveals that the elastic and the inelastic scattering processes have the same relative phase.
Here $\omega_{in} = 170$ a.u. , $E_{\omega_{in}} = 10.7 $ a.u. , $E_{2\omega_{in}} = 0.535 $ a.u. , $t_{wid} = 3$ a.u. , $Z = 4$ and $a^2 = 0.1$ a.u. with BE = 4.805 a.u.
}
\end{figure}

We examine if the intrinsic phase difference ($\delta$) depends on the pulse width ($t_{wid}$). We increase the $t_{wid}$ to 3 a.u., which is 30 times the pulse width used in Fig.~\ref{interference_3401C_135}. There are computational challenges associated with this long calculation, so a 2D calculation is performed instead. For a detailed discussion on a 2D treatment of nonlinear Compton scattering, see Ref.~\cite{NLCPRA_1}. For this calculation, fields with different intensities are chosen to illustrate the difference in their scattering profile. The electric field amplitudes  for $\omega_{in}$ field ($E_{\omega_{in}}$) and $2\omega_{in}$ field ($E_{2\omega_{in}}$), are chosen to be 10.7 a.u.  and 0.535 a.u. respectively. The results of the 2D calculation reveal an intrinsic phase difference ($\delta$) that is 0 or $\pi$ depending on the scattering angle. Increasing the pulse width decreases the bandwidth of the incoming pulse. The smaller bandwidth reveals two scattering mechanisms for the incoming photons (see Fig.~\ref{Fig_largetwid} ). In Fig.~\ref{Fig_largetwid}, the first peak $k\sim 2.41$ a.u. centered around the Compton scattered momentum for the $2\omega_{in}$ field, arises from the inelastic scattering of the incoming photons by the electron. The second peak $k\sim 2.48$ a.u., sharply centered around the incoming wavenumber of the $2\omega_{in}$ field, arises from the elastic scattering of the incoming photons by the electron, leaving the electron in the ground state. Note that the contribution of this elastic scattering peak to the overall area under the curve is small (see Fig.~\ref{Fig_largetwid}). The figure (Fig.~\ref{Fig_largetwid}) shows that the intrinsic phase difference between the Compton and non-linear Compton scattering is the same for the elastic and the inelastic process.

The effect of polarization directions can be interesting. For a single incoming field, there are 4 possible orientations based on the direction of the incoming field's polarization($\epsilon_{in}$) and momentum($k_{in}$) and the polarization~($\epsilon$) and the momentum~($k$) of the outgoing fields. We choose the 4 cases in the following manner:
\begin{enumerate}
\item Case 1: $\boldsymbol{\epsilon_{in}} = \hat{y}$, $\boldsymbol{k_{in}} = \hat{x}$, $\boldsymbol{\epsilon} = -\hat{x}~sin~\theta  + \hat{y} ~ cos~\theta $ and $\boldsymbol{k} = \hat{x} ~ cos~\theta ~  + \hat{y} ~ sin~\theta $ ; i.e. the initial and final polarizations \textit{in} the scattering plane. An abbreviated notation of in-in is used to denote this case.
\item Case 2: $\boldsymbol{\epsilon_{in}} = \hat{y}$, $\boldsymbol{k_{in}} = \hat{x}$, $\boldsymbol{\epsilon} = \hat{z}$ and $\boldsymbol{k} = \hat{x}~ cos~\theta  + \hat{y}~sin~\theta $ ; i.e. the initial polarization \textit{in} the scattering plane but the final polarization \textit{perpendicular} to the scattering plane. An abbreviated notation of in-out is used to denote this case.
\item Case 3: $\boldsymbol{\epsilon_{in}} = \hat{z}$, $\boldsymbol{k_{in}} = \hat{x}$, $\boldsymbol{\epsilon} = -\hat{x}~sin~\theta  + \hat{y} ~ cos~ \theta  $ and $\boldsymbol{k} = \hat{x} ~ cos~\theta + \hat{y}~sin~\theta $ ; i.e. the initial polarization \textit{perpendicular} to the scattering plane but the final polarization \textit{in} the scattering plane. An abbreviated notation of out-in is used to denote this case.
\item Case 4: $\boldsymbol{\epsilon_{in}} = \hat{z}$, $\boldsymbol{k_{in}} = \hat{x}$, $\boldsymbol{\epsilon} = \hat{z}$ and  $\boldsymbol{k} = \hat{x}~cos~\theta  + \hat{y}~sin~\theta$ ; i.e. the initial and final polarizations \textit{perpendicular} to the scattering plane. An abbreviated notation of out-out is used to denote this case.
\end{enumerate}

Note that for interference to be possible, the scattered photons from both Compton and nonlinear Compton should have the same final polarization and momentum vector. Also, for nonlinear Compton scattering, from Eq.~(\ref{brownkibbleformula}) only Case 1~(in-in) and Case 3~(out-in) are expected to yield non-zero scattering probabilities (more on this point below). Therefore based on these two requirements, only the interference cases with Compton scattering corresponding to Case 1~(in-in) or Case 3~(out-in) and nonlinear Compton scattering pertaining to Case 1~(in-in) or Case 3~(out-in) are significant. The results presented so far (Figs.~\ref{interference_3401C_135} - \ref{Fig_largetwid} ) correspond to Case 1~(in-in) for both linear and nonlinear Compton scattering. While we expect this to be the dominant case from Eq.~(\ref{Kleinnishinaformula}) and Eq.~(\ref{brownkibbleformula}), the other cases become relevant when one considers the interference effects from crossed polarizations or from unpolarized photons. From here on, an abbreviated notation of, for example, (in-in, out-in) refers to the interference when the Compton scattering pertains to Case 1 (in-in) and nonlinear Compton scattering pertains to Case 3~(out-in). 

The nonlinear Compton scattering probability pertaining to Case 3~(out-in), is expected to be approximately the same size as Case 1~(in-in) from Eq.(\ref{brownkibbleformula}). For $\theta \in [0,180]$ the differential cross section for Case 1~(in-in) has zeroes at $\theta = 0\degree,~ \sim 75\degree$ and $180\degree$ but Case 3~(out-in) has zeroes only at $\theta = 0\degree$ and $180\degree$. We examine interference for the case (in-in, out-in) for the same set of parameters as Fig.~\ref{Bnormalized}. The intrinsic phase difference $\delta$ is found to have a similar dependence on the scattering angle as that of (in-in, in-in) interference in that, it switches between 0 or $\pi$ every time the scattering angle crosses a zero of the differential cross section of Compton or nonlinear Compton scattering~\cite{KB}. For a given scattering angle, the calculations with $Z=2$ and $Z=4$ show a small difference in the differential cross sections of about $\sim 0.014~\% $. But, they both still have the same intrinsic phase $\delta$.

The polarization directions are now chosen to be Case 3~(out-in) for Compton scattering and Case 1~(in-in) for nonlinear Compton scattering. This is an interesting case because for such an arrangement we expect the nonlinear Compton signal to be comparable to the Compton signal. This arises out of the interplay of two factors. First, for the chosen intensity ($E = 107$ a.u.) nonlinear Compton signal from the incoming field is less than that of the Compton signal. Second, the Compton scattering for this arrangement is suppressed because of the choice of polarization (Case 3~(out-in)). The results of the calculation reveal that the size of the Compton scattered wave function (${\psi}_{1,2\omega_{in}}^{(1)}$) is in fact comparable to that from nonlinear Compton scattering (${\psi}_{2,\omega_{in}}^{(1)}$). However, no interference occurs because the scattered wave function for Compton scattering and that for nonlinear Compton scattering are found to be orthogonal to each other. This is found to be a consequence of the fact that the Compton scattered wave function is anti-symmetric in the $\hat{z}$ direction but the nonlinear Compton scattered wave function is symmetric along the same direction. These symmetries can be deduced from the form of the perturbation equations.

Consider the equations from the perturbative approach (Sec.~\ref{perturbative_approach}) keeping in mind our choice of Case 3~(out-in) for Compton scattering and Case 1~(in-in) for nonlinear Compton scattering. The symmetric or the anti-symmetric nature of the scattered wave function can be understood by tracking the effect of the source terms involved. The Hamiltonian for the atomic electron ($H_a$) is parity-symmetric and the starting wave function (${\psi}_0^{(0)}$) being the ground state of the electron is symmetric. The source terms in all the equations from Eq.~(\ref{homogeneous_zeroth}) - (\ref{homogeneous_second}) and from Eq.~(\ref{Inhomogeneous_zeroth}) - (\ref{Inhomogeneous_second}), have definite parity in the z-direction. Therefore, the wave functions of different perturbative order also have definite parity in the z-direction, since the homogeneous part of the equations preserves parity. One can track the changes in the parity of the ground state wave function of the electron (${\psi}_0^{(0)}$) from each source term in the first order perturbative treatment for Compton scattering and the second order treatment for nonlinear Compton scattering respectively. Such an analysis can be used to determine that the scattered wave function for Compton scattering (${\psi}_{1,2\omega_{in}}^{(1)}$) is anti-symmetric in the z-direction for Case 3~(out-in) and the scattered wave function for nonlinear Compton scattering (${\psi}_{2,\omega_{in}}^{(1)}$) is symmetric in the z-direction for Case 1~(in-in). Equivalently, instead of parity arguments this can also be understood through the number of insertions of the electric field in the source terms of perturbative equations for Compton and nonlinear Compton scattering.

For the case of polarizations (out-in, out-in), we find that similar to the previous case there is no interference. Again, one can use a similar approach using perturbative equations to deduce that the Compton scattered wave function for Case 3~(out-in) is anti-symmetric and the nonlinear Compton scattered wave function for Case 3~(out-in) is symmetric in $z$.

Upon exploring other cases for polarization, we find that for some cases we expect a zero scattering probability for nonlinear Compton scattering~[Eq.~(\ref{brownkibbleformula})]. Consider the case when the polarization of the scattered photons for nonlinear Compton pertains to Case 4~(out-out). From Eq.~(\ref{brownkibbleformula}), one would expect a zero scattering probability. The calculations however reveal a non-zero but small scattering probability. It is small when compared to nonlinear Compton scattering probability of Case 1~(in-in). This expectation of zero scattering probability is a consequence of the assumption~\cite{KB} that the electron is initially at rest. In our calculations, this is not the case because of the Compton profile of the ground state electron. The effect of the Compton profile on the nonlinear Compton scattering probability can be studied by examining the scattering from a free electron modelled by a Gaussian wave packet. Decreasing the spatial width of the initial free electron wave packet~(${\psi}_0^{(0)}$) along the z-direction widens the initial z-momentum distribution of the electron. Therefore, from Eq.~(\ref{Exact_eqn_psi1}) the scattering probability is expected to be proportional to the inverse square of the spatial width in the z-direction of the wave packet. Numerical calculations for nonlinear Compton scattering for Case 4~(out-out), confirms this behaviour. Therefore it is clear, that the non-zero scattering probability for nonlinear Compton scattering for Case 4~(out-out) is the effect of the Compton profile.

\section{Conclusion and Summary}
The interference between Compton scattering and nonlinear Compton scattering from two incoming fields was examined. To understand the phase shifts involved, a first order perturbative approach in the incoming classical field was used to describe Compton scattering and a second order perturbative approach was used to describe nonlinear Compton scattering. The regimes where the approach is valid was analyzed by comparing it with a previously developed approach that was non-perturbative in the classical field~\cite{NLCPRA_1}. The effect of the polarization of the incoming and outgoing photons on the the interference was studied. For interference to exist, the scattered wave vector and polarization of the scattered wave for both Compton scattering and nonlinear Compton scattering have to be the same. As a reminder, an abbreviated notation of say (in-in, out-in) denotes that for Compton scattering the incoming and outgoing photon polarizations are both in the scattering plane and for nonlinear Compton scattering the incoming photon polarization is out of the scattering plane and the outgoing photon polarization is in the scattering plane. 

For the case of (in-in, in-in), the results of the numerical calculation shows that the phase shift between Compton scattering and nonlinear Compton scattering was either 0 or $\pi$, switching between the two, every time the scattering angle crosses a zero in the differential cross section of Compton or nonlinear Compton scattering~\cite{KB}. A similar behaviour for the intrinsic phase difference is found for (in-in, out-in) interference.

For both (out-in, in-in) interference and (out-in, out-in) interference, it was found that the scattered wave functions for Compton and nonlinear Compton scattering were orthogonal to each other. Therefore, no interference was found to exist. 

These results can help with two common experimental challenges in nonlinear Compton scattering. First, the interference could be used to detect the small nonlinear Compton scattering signal. Second, the interference could be used to distinguish the Compton scattering noise originating from the second-harmonic of the XFEL and the nonlinear Compton scattering signal from the fundamental harmonic of the XFEL. 

\section{Acknowledgements}
This work was supported by the U.S. Department of Energy, Office of
Science, Basic Energy Sciences, under Award No. DE-SC0012193. We express our gratitude to D.A. Reis for fruitful discussions on nonlinear Compton scattering and the experimental challenges involved. We are grateful to Y.N. Pushkar for discussions on X-ray scattering. A.V thanks S. Vaidya for helpful discussions on non-homogeneous partial differential equations.

\bibliography{References.bib}

\end{document}